%% file: muonic_forces_LHC_arXiv_v1.tex
\documentclass[10pt,aps,prd,nofootinbib,superscriptaddress,
  twocolumn,preprintnumbers,balancelastpage]{revtex4}
\PassOptionsToPackage{slash}{url}
\usepackage{graphicx,epsfig,psfrag,amssymb,hyperref}
\usepackage{multirow}
\usepackage{color,graphicx,epsfig,psfrag,amsmath,empheq}
\usepackage{bm}
\usepackage{mathrsfs,amsfonts,color}
\usepackage{slashed}
\usepackage{hepunits}
\usepackage{color}
\usepackage{enumitem}
\usepackage{booktabs}
\usepackage{xspace}
\usepackage[utf8]{inputenc}

\AtBeginDocument{
\heavyrulewidth=.08em
\lightrulewidth=.05em
\cmidrulewidth=.03em
\belowrulesep=.65ex
\belowbottomsep=0pt
\aboverulesep=.4ex
\abovetopsep=0pt
\cmidrulesep=\doublerulesep
\cmidrulekern=.5em
\defaultaddspace=.5em
}

\newcommand{\abs}[1]{\left\lvert#1\right\rvert}

\newcommand{\nb}{\ensuremath{\mathrm{nb}}\xspace}
\newcommand{\ifb}{\ensuremath{\mathrm{fb}^{-1}}\xspace}
\newcommand{\iab}{\ensuremath{\mathrm{ab}^{-1}}\xspace}

\renewcommand{\GeV}{\ensuremath{\mathrm{Ge\mkern-1.5muV}}\xspace}
\renewcommand{\MeV}{\ensuremath{\mathrm{Me\mkern-1.5muV}}\xspace}

\newcommand{\g}{\ensuremath{\mathrm{g}}\xspace}

\renewcommand{\eqref}[1]{Eq.~(\ref{eq:#1})}

\newcommand{\figref}[1]{Fig.~\ref{fig:#1}}

\newcommand{\nn}{\nonumber}

\newcommand{\cP}{\mathcal{P}}
\newcommand{\cL}{\mathcal{L}}

\newcommand{\cA}{\mathcal{A}}
\newcommand{\cM}{\mathcal{M}}
\newcommand{\cR}{\mathcal{R}}
\newcommand{\cO}{\mathcal{O}}

\newcommand{\pMS}{p_{\rm MS}}
\newcommand{\pID}{p_{\rm ID}}
\newcommand{\pME}{p_{\rm ME}}
\newcommand{\pin}{p_{\rm in}}
\newcommand{\pout}{p_{\rm out}}
\newcommand{\Ein}{E_{\rm in}}
\newcommand{\Eout}{E_{\rm out}}
\newcommand{\ECAL}{E_{\rm{cal}}}
\newcommand{\LFT}{\mathcal{L}^{\rm{int}}_{\rm{FT}}}
\newcommand{\LLHC}{\mathcal{L}^{\rm{int}}_{\rm{LHC}}}
\newcommand{\pT}{p_T}
\newcommand{\NMFC}{N_X}
\newcommand{\NSM}{N_{\rm SM}}
\newcommand{\mmm}{$\rm{M}^3$}

\newcommand{\secvspace}{\vspace{5mm}}
\newcommand{\subsecvspace}{\vspace{3mm}}

\begin{document}

%%%%%%%%%%%%%%%%%%%%%%%%%%%%%%%%%%%%%%%%%%%%%%%%
\title{Searching for muonic forces with the ATLAS detector}
%%%%%%%%%%%%%%%%%%%%%%%%%%%%%%%%%%%%%%%%%%%%%%%%

%%%%%%%%%%%%%%%%%%%%%%%%%%%%%%%%%%%%%%%%%%%%%%%%
\author{Iftah Galon}
\email{iftah.galon@physics.rutgers.edu}
\affiliation{NHETC, Dept.~of Physics and Astronomy, Rutgers University, Piscataway, NJ 08854 USA}

\author{Enrique Kajamovitz}
\email{enrique@physics.technion.ac.il}
\affiliation{Department of Physics, Technion, Haifa 32000, Israel}

\author{David Shih}
\email{dshih@physics.rutgers.edu}
\affiliation{NHETC, Dept.~of Physics and Astronomy, Rutgers University, Piscataway, NJ 08854 USA}

\author{Yotam Soreq}
\email{yotam.soreq@cern.ch}
\affiliation{Theoretical Physics Department, CERN, CH-1211 Geneva 23, Switzerland}
\affiliation{Department of Physics, Technion, Haifa 32000, Israel}

\author{Shlomit Tarem}
\email{shlomit.tarem@cern.ch}
\affiliation{Department of Physics, Technion, Haifa 32000, Israel}
%%%%%%%%%%%%%%%%%%%%%%%%%%%%%%%%%%%%%%%%%%%%%%%%

%%%%%%%%%%%%%%%%%%%%%%%%%%%%%%%%%%%%%%%%%%%%%%%%
\begin{abstract}
The LHC copiously produces muons via different processes, and the muon sample will be large at the high-luminosity LHC~(HL-LHC). 
In this work we propose to leverage this large muon sample and utilize the HL-LHC as a muon fixed-target experiment, with the ATLAS calorimeter as the target. 
We consider a novel analysis for the ATLAS detector, which takes advantage of the two independent muon momentum measurements by the inner detector and the muon system. 
We show that a comparison of the two measurements, before and after the calorimeters, can probe new force carriers that are coupled to muons and escape detection. 
The proposed analysis, based on muon samples from $W$ and $Z$ decays only, has a comparable reach to other proposals. In particular, it can explore the part of parameter-space that could explain the muon $g-2$ anomaly.
\end{abstract}
%%%%%%%%%%%%%%%%%%%%%%%%%%%%%%%%%%%%%%%%%%%%%%%%

\preprint{CERN-TH-2019-097}

\maketitle

%%%%%%%%%%%%%%%%%%%%%%%%%%%%%%%%%%%%%%%%%%%%%%%%
\textbf{\textit{Introduction}} \newline
%%%%%%%%%%%%%%%%%%%%%%%%%%%%%%%%%%%%%%%%%%%%%%%%
The Standard Model~(SM) of particle physics has been successful in describing the known elementary particles and their interactions and is directly tested by experiments up to the TeV scale. 
Nevertheless, the SM is not a complete description of Nature, and should be augmented by new physics~(NP) degrees of freedom which account for neutrino oscillations, dark matter~(DM), and the matter/anti-matter asymmetry of the Universe.

One possible manifestation of NP are new particles with masses in the MeV--to--GeV range and suppressed couplings to the SM.
Such new particles could be part or all of DM, or act as mediators to a dark sector.
Muonic Force Carrier (MFC) mediators, $X$, are particularly interesting.  
These mediators have flavor-specific couplings~\cite{Batell:2016ove,Chen:2017awl,Batell:2017kty,Marsicano:2018vin}, couple to the SM only through muons and may decay predominantly to DM. 
MFCs potentially explain inconsistencies in low-energy observations such as the anomalous magnetic dipole moment of the muon~\cite{Bennett:2006fi,Beringer:1900zz}, and the possible anomaly in the measurement of the proton radius in muonic hydrogen~\cite{Pohl:2013yb,Hill:2016bjv}.  

Existing constraints on the existence of dark sector mediators are predominantly derived from beam-dump, fixed-target, or collider experiments~\cite{Alexander:2016aln,Battaglieri:2017aum,Beacham:2019nyx}. 
The constraints are weaker for models where the mediator couplings to electrons or protons are suppressed. 
Specifically, MFC mediators are only weakly constrained~\cite{Chen:2017awl}.  Models where $m_X>2m_\mu$ and $X$ dominantly decays back to $\mu^+\mu^-$ are constrained by the BaBar analysis~~\cite{TheBABAR:2016rlg}. 
Data from rare $B$ decays also constrain MFC mediators~\cite{Chala:2019vzu}, but with larger model dependency.

Recent studies suggest that MFCs could be searched for in muon-fixed-target experiments~\cite{Chen:2017awl,Kahn:2018cqs,Chen:2018vkr}, in kaon decays at the NA62 experiment~\cite{Krnjaic:2019rsv} or in Belle~II~\cite{Jho:2019cxq}. 
MFC production in muon-target interactions would register as a momentum difference between the incoming and outgoing muons which is not accounted for by the energy deposition in the (instrumented) target. 
Such dedicated apparatuses may be available at CERN by running the NA64 experiment with a muon beam~\cite{Chen:2017awl,Gninenko:2019qiv}, and at FermiLab, by leveraging the muon beam-line of the Muon~$(g-2)$ experiment~\cite{Grange:2015fou}, and constructing the \mmm~apparatus~\cite{Kahn:2018cqs}.

\begin{figure}[t]
\includegraphics[width=0.5\textwidth]{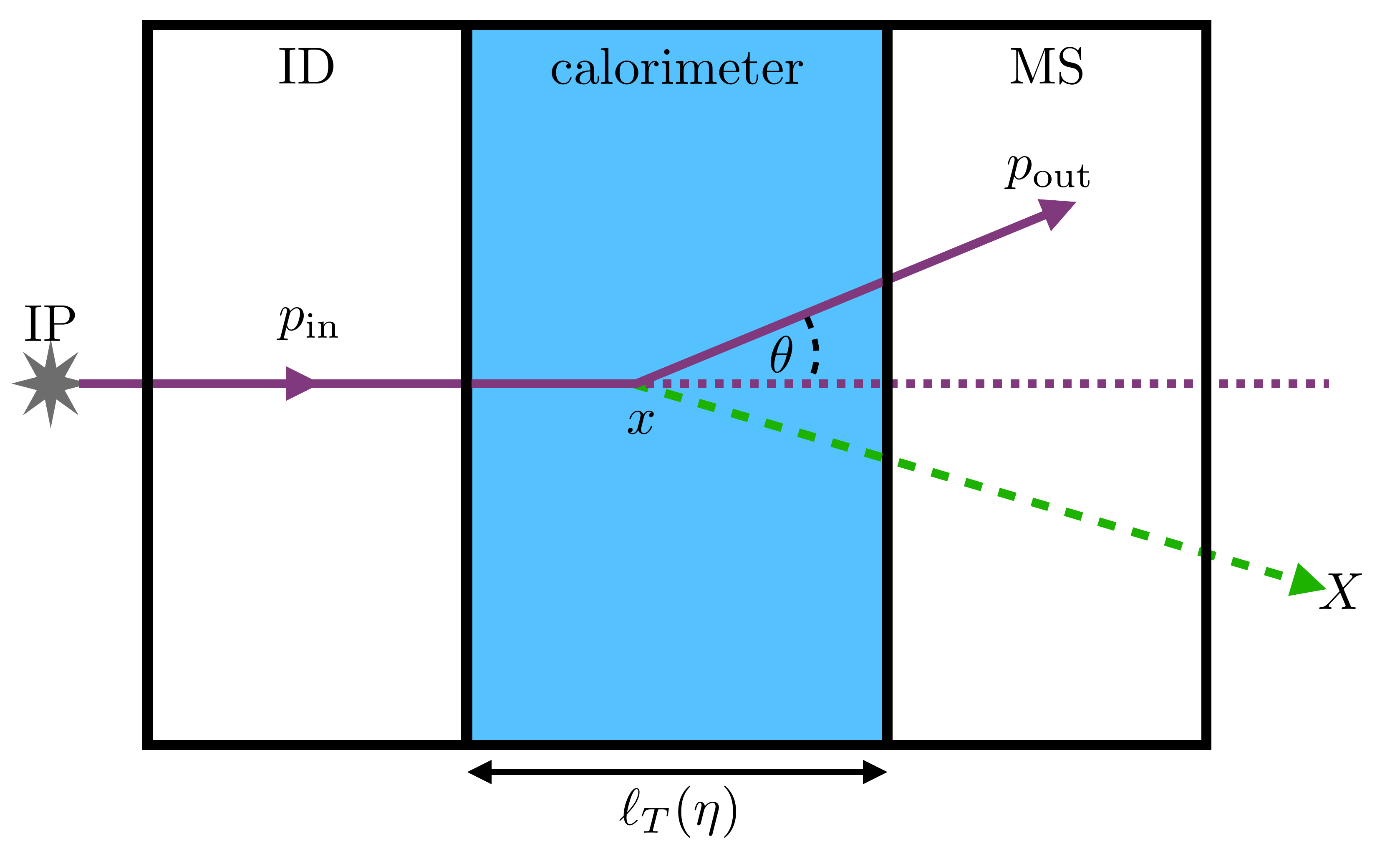}
\caption{An illustration of the proposed measurement - the ATLAS detector as muon fixed-target experiment.
}
\label{fig:illust} 
\end{figure}

In this Letter, we propose to utilize the ATLAS detector as a muon fixed-target experiment, which is sensitive to the missing muon momentum signature and therefore probes MFCs.
The calorimeters serve as an instrumented target, and the Inner Detector~(ID), and Muon System~(MS) provide
independent muon momentum measurements before and after the target, as illustrated in \figref{illust}. It is important that in ATLAS there is no un-instrumented material between the calorimeter and the MS. This ensures that an accurate measurement of the missing muon  momentum signature is possible.
%%%%%%%%%%%%%%%%%%%%%%%%%%%%%%%%%%%%%%%%%%%%%%%%
%%%%%%%%%%%%%%%%%%%%%%%%%%%%%%%%%%%%%%%%%%%%%%%%

Using muons from $Z$ and $W$ decays, in total $\cO(10^{10})$ muons on target at the high-luminosity LHC~(HL-LHC), we estimate that the proposed analysis is sensitive to MFC masses in the MeV--GeV range with couplings as low as $g_X \sim 10^{-4}-10^{-3}$. 
This MFC parameter-space includes the region which is relevant to account for the $(g-2)_\mu$ anomaly and is comparable to other proposals such as \mmm~phase~1 and NA62.

\secvspace

%%%%%%%%%%%%%%%%%%%%%%%%%%%%%%%%%%%%%%%%%%%%%%%%
\textbf{\textit{Benchmark models}} \newline
%%%%%%%%%%%%%%%%%%%%%%%%%%%%%%%%%%%%%%%%%%%%%%%%
We use as MFC benchmarks scalar or vector mediators, i.e. $X=S$ or $V$~\cite{Chen:2017awl,Kahn:2018cqs,Chen:2018vkr,Krnjaic:2019rsv,Jho:2019cxq}.
The effective interaction Lagrangians are given by
\begin{align}
	\label{eq:LSV}
	\cL_{V} 
= 	g_V V_\alpha \bar\mu\gamma^\alpha\mu \, ,
	\qquad\qquad
	\cL_{S}
=	g_S S \bar\mu \, \mu \,,
\end{align}
where we have omitted the mass and kinetic terms.
We assume that $X$ is a mediator to a dark sector which predominantly decays into undetected particles, i.e. implicitly assume large couplings of $X$ to sufficiently light dark sector constituents. 
Alternatively, $X$ could be sufficiently long-lived so as to escape detection. 

The effective interaction in \eqref{LSV} can be UV completed. 
For example, the vector interaction may arise in a broken gauged $L_\mu-L_\tau$ gauge theory~\cite{He:1991qd}.
The scalar interaction can be a result of interactions with heavy leptons, which are integrated out~\cite{Krnjaic:2019rsv}.

The simplified models in \eqref{LSV} are subject to existing constraints depending on their UV completion.  
Here we consider 
(a)~the muon magnetic moments~\cite{Bennett:2006fi,Beringer:1900zz}, $(g-2)_\mu$;
and (b)~from CHARM-II $\mu$-trident~\cite{Geiregat:1990gz,Altmannshofer:2014pba}. 
The bound from $(g-2)_\mu$ can be avoided in models where different contributions to the loop cancel each other, for example, scalar and vector against pseudo-scalar and axial-vector, e.g. see~\cite{Jegerlehner:2009ry}.
The $\mu$-trident bound is valid only for vector mediators with left-handed coupling to the muon. 

\secvspace

%%%%%%%%%%%%%%%%%%%%%%%%%%%%%%%%%%%%%%%%%%%%%%%%
\textbf{\textit{ATLAS as a muon fixed-target experiment}} \newline
%%%%%%%%%%%%%%%%%%%%%%%%%%%%%%%%%%%%%%%%%%%%%%%%
Muons produced at the ATLAS interaction point~(IP) traverse the entire detector, leaving signals in the ID, calorimeters, and MS.
ATLAS muon reconstruction~\cite{Aad:2016jkr} is first performed independently in the ID and MS, with each detector sub-system providing muon spatial location and transverse-momentum ($\pT$) measurements. 
Subsequently the ID and MS information is combined with the calorimeter measurement to form the muon tracks which are used in physics analyses. 

ATLAS defines four muon types, according to the details of the combination procedure, of which two are relevant for this work
\begin{itemize}
\item A combined~(CB) muon track is formed from independent tracks in the ID and MS, with a global refit that uses the hits from both sub-detectors. 
\item Extrapolated~(ME) muons have trajectories reconstructed based only on the MS track and a loose requirement on compatibility with originating from the IP. 
The muon track parameters are defined at the IP, and take into account the muon energy loss estimation,  $\pME \approx \pMS + \ECAL$. 
The latter estimate combines the calorimeter measurement with a detailed analytic parameterization of the average energy loss,
a method which yields a precision of $\sim30\,\MeV$ for $50\,\GeV$ muons. ME muons are ideal candidates for an ATLAS search of the MFC signal. 
\end{itemize}

MFC production in the muon-target interaction manifests as
\begin{align}
	\label{eq:discriminant}
	\pMS + \ECAL  - \pID  <  0\, ;
\end{align}
a difference between $\pID$ and $\pMS$ that is not compensated by $\ECAL$. 
We define an observable which combines the ID momentum measurement, $\pID$, with, $\pME$, the reconstructed momentum of an ME type muon, 
\begin{align}
	\label{eq:rho}
	\rho \equiv 
	\frac{ \pME  - \pID }{ \pID } 
	\approx
	\frac{ \pout  - \pin }{\pin  }  \, ,
\end{align}
where, up to resolution effects, we identify the incoming\,(outgoing) muon momentum with respect to the target, $\pin\,(\pout)$, with $\pID\,(\pME)$.

The tag-and-probe method with $Z\to\mu\mu$, where one muon is reconstructed as a CB muon (tag) and the second may be a ME muon (probe), provides a high-purity muon sample~\cite{Aad:2016jkr} with loose selection on the probe muon that can be used to search for MFCs. 
We foresee that with careful analysis it will be possible to also use $W\to\mu\nu$ decays.

\secvspace

%%%%%%%%%%%%%%%%%%%%%%%%%%%%%%%%%%%%%%%%%%%%%%%%
\textbf{\textit{Sensitivity estimation}} \newline
%%%%%%%%%%%%%%%%%%%%%%%%%%%%%%%%%%%%%%%%%%%%%%%%
Next, we estimate the sensitivity of the proposed analysis to probe MFCs, which are described by the interactions in \eqref{LSV}. 
Throughout we will normalize our projections to the expected integrated luminosity of the HL-LHC, $\LLHC = 3\,\iab$.

\subsecvspace

%%%%%%%%%%%%%%%%%%%%%%%%%%%%%%%%%%%%%%%%%%%%%%%%
\textit{Muon-target luminosity and MFC production rate}\newline
%%%%%%%%%%%%%%%%%%%%%%%%%%%%%%%%%%%%%%%%%%%%%%%%
For minimally ionizing particles such as muons, the fixed-target effective luminosity is given by:
\begin{align}
	\label{eq:fixed_target_lumi}
	\LFT 
\!= &	 N_\mu  \frac{ \rho_T}{A \,m_0}  \Delta x  \nonumber \\
=&	\left( \frac{\cL^{\rm int}_{\rm LHC} }{3\,\iab} \right)\!	
	\left( \frac{\sigma^{\rm fid}_{\rm prod} }{\nb} \right)\!
	\left( \frac{63}{A} \right) \nonumber\\
&	\times\left(\frac{\rho_T}{8.96\,\g/\cm^3}\right)\!
	\left(\frac{\Delta x}{253\,\cm}\right)\!
	65\,\nb^{-1} \, ,
\end{align}
where we have treated the ATLAS calorimeter\footnote{The ATLAS calorimeter ranges up to $\Delta x = 144\,X_0$, where $X_0=1.757\,\cm$ is the radiation length of electrons in iron.} as a thin target, and assumed a single material composition,\footnote{While the calorimeter targets are comprised of various materials, the variation in the event yield is small, and the effect on the experimental sensitivity is negligible.} with density $\rho_T$, mass number $A$, length $\Delta x$ and $N_\mu = \LLHC \sigma^{\rm fid}_{\rm prod} $ incoming muons. 
Here, $m_0= 1.661 \times 10^{-24}\,\g$, and $\sigma^{\rm fid}_{\rm prod}$ is the (process dependent) cross section for muon production at the LHC, within the ATLAS fiducial volume.
In this work, we assume a $^{63}_{29}\rm{Cu}$ target, which corresponds to  $A=63$ and $\rho_T =8.96\,\g/\cm^3$ in \eqref{fixed_target_lumi}.

We estimate the MFC signal production rate following the schematics in \figref{illust}.
A muon originating from the IP with momentum $\pin$ and direction $\eta$ interacts at a point $x$ in the material target of length $\ell_T(\eta)$, and produces a MFC, and an outgoing muon of momentum $\pout$ which travels in an angle $\theta$ relative to the incoming muon direction. 
The expected number of produced MFCs within the detector acceptance, $\cA$ (which includes the MFC target interaction) is:\footnote{Note that we have assumed that the detector is cylindrically symmetric, however, $\phi$ integration can be straightforwardly incorporated to account for possible inhomogeneities.}
\begin{align}
	\label{eq:yield}
	\NMFC
&=\,	\LFT \,  \sigma_T
	 \int \! dp_{\rm in} \int  d\eta \,\,  \mathcal{A} \times \cP_{\rm in}(\pin, \eta) \, ,
\end{align}
where $\sigma_T$ is the fixed-target MFC production cross section, $\mu \, T \to \mu \, T  \, X $, and $\cP_{\rm in}(\pin,\eta)$, 
is the incoming muon double-differential distribution:
\begin{align}
	\label{eq:Psig}
	\cP_{\rm in}(\pin,\eta)
	\equiv
	\frac{1}{\sigma^{\rm fid}_{\rm prod}} \frac{d^2\sigma^{\rm fid}_{\rm prod}}{d \pin\, d \eta} \, .
\end{align}
In addition, below we will be interested in $d\NMFC/d\rho$, which is straightforward to derive from \eqref{yield}. 
For further details on the $\NMFC$ estimation see the Supplemental Materials.

\subsecvspace

%%%%%%%%%%%%%%%%%%%%%%%%%%%%%%%%%%%%%%%%%%%%%%%%
\textit{Signal yields}\newline
%%%%%%%%%%%%%%%%%%%%%%%%%%%%%%%%%%%%%%%%%%%%%%%%
We use Monte-Carlo~(MC) simulations to estimate the signal yield and its $\rho$ distribution. 
We consider muons from $Z$ and $W$ decays because they provide high-purity muon samples.\footnote{
It would also be interesting to consider other sources of muons at the LHC, {\it e.g.} from heavy flavor or $J/\psi$'s decays. 
There are potentially many more of these muons but they would of course be much more difficult to use for an MFC search due to the large background and lower $\pT$'s.
}
In the MC sample, events are selected following the criteria in Ref.~\cite{Aad:2016naf}. In the selected events, only muons in the barrel ($0.1<\abs{\eta}<1.05$) are used, since the ID momentum resolution in the barrel is better, and the depth of the calorimeter is approximately constant.
In addition, we require a weak ID to MS angular matching requirement of $\theta<0.1$ to account for the loose IP matching of ME muons. 

The incoming muon momentum spectrum, $\cP_{\rm fid}(\pin)\equiv \int d\eta \, \cP_{\rm in}(\pin,\eta)$, is obtained from a {\sc MadGraph\,5 v\,2.6.1} simulation~\cite{Alwall:2014hca, deAquino:2011ub} of the hard process, with up to two additional jets, and interfaced with {\sc Pythia\,8.230}~\cite{Sjostrand:2007gs} for showering and hadronization. 
We apply the MLM jet-matching scheme~\cite{Mangano:2006rw} to combine the different samples, and we use the {\sc Delphes\,3.4.1}~\cite{deFavereau:2013fsa} detector simulation with the standard  ATLAS card. 
The momentum spectrum of the selected muons is taken at the truth level of MC generation and smeared based on the published ATLAS muon momentum resolutions ($\sigma_{\rm in}=0.015\,{\rm MeV}+ 3 \times 10^{-7}\pin$ and $\sigma_{\rm out}=0.05\,\pout$, see~\cite{Aad:2016jkr}).

\begin{figure}[t]
\includegraphics[width=0.5\textwidth]{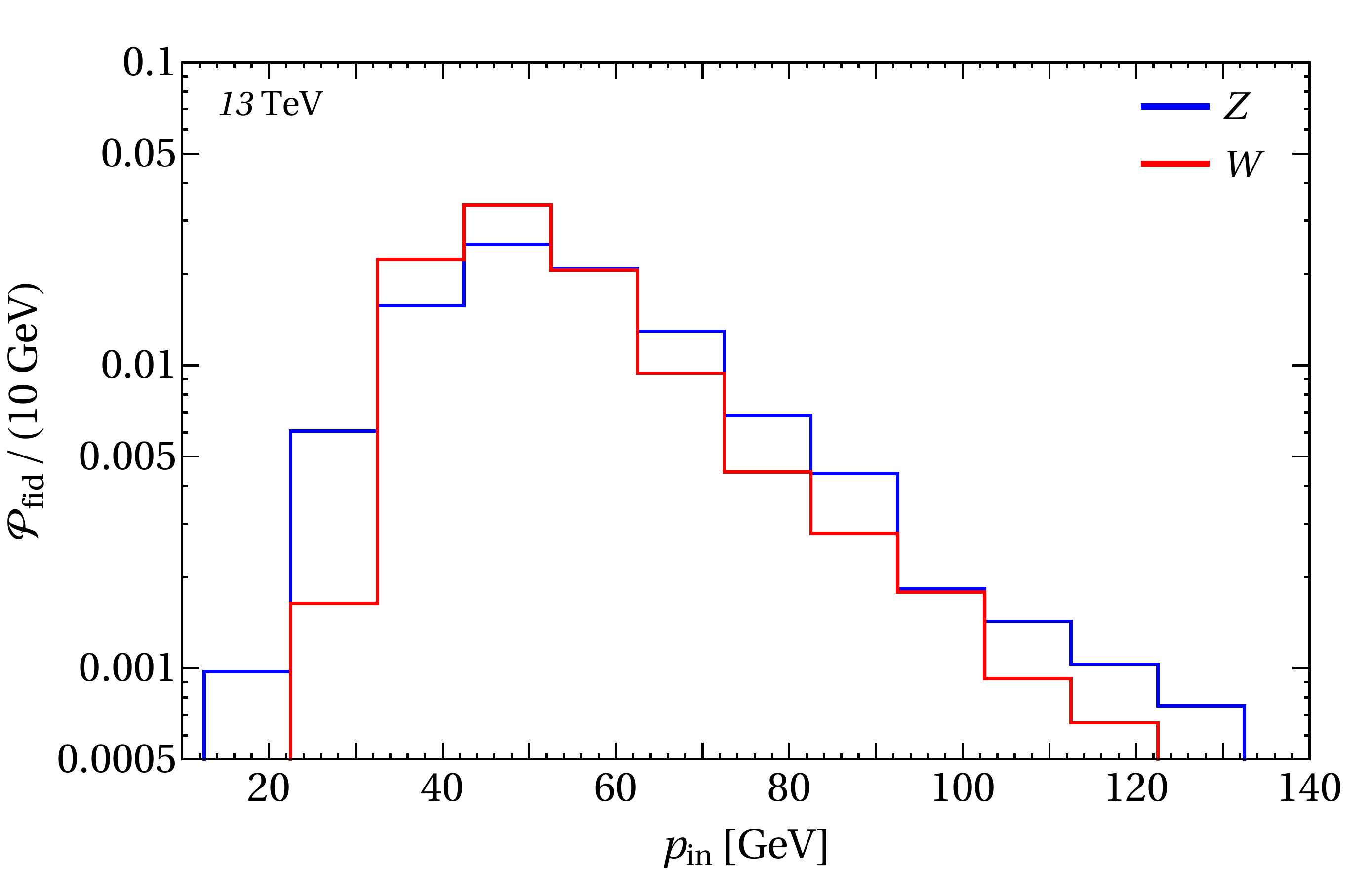}
\caption{
The differential momentum distributions of muons in our fiducial region, $\cP_{\rm fid}(\pin)$, from $Z$ (blue) and $W$ (red) decays.
The event selection criteria follow those in~\cite{Aad:2016naf}, 
The plots are normalized to $1$.
}
\label{fig:Zmuons}
\end{figure}

We normalize the muon production rate to match the ATLAS result of $Z\to\mu\mu$ and $W\to\mu\nu$~\cite{Aad:2016naf,Aaboud:2017hbk, Sirunyan:2018cpw}, and find  
\begin{align}
	\label{eq:fidZ}
	\sigma^{\rm fid}_Z
=&  	\epsilon^{\rm eff}_Z \, \sigma^{\rm ATLAS}_{pp\to Z \to \mu\mu} = 0.39  \, \nb \, ,  \\
	\label{eq:fidW}
	\sigma^{\rm fid}_W
=& 	\epsilon^{\rm eff}_W \, \sigma^{\rm ATLAS}_{pp\to W \to \mu\nu} = 3.5  \, \nb \, ,
\end{align}
where $\sigma^{\rm ATLAS}_{pp\to Z \to \mu\mu}\,(\sigma^{\rm ATLAS}_{pp\to W \to \mu\nu})=0.78\,(8.0)\,\nb$~\cite{Aad:2016naf}, and  $\epsilon_Z^{\rm eff} \, (\epsilon_W^{\rm eff}) = 0.50\,(0.44)$ is the efficiency factor relating the ATLAS cuts in Ref.~\cite{Aad:2016naf} with our selection of the barrel as the fiducial volume.
The resulting momentum spectra are shown in \figref{Zmuons}.
%%%%%%%%%%%%%%%%%%%%%%%%%%%%%%%%%%%%%%
%%%%%%%%%%%%%%%%%%%%%%%%%%%%%%%%%%%%%%

The $\mu T \to \mu T X$ process is simulated in {\sc MadGraph\,5}, including the target nuclear-atomic form factor by modifying the photon-target vertex, see Supplemental Material for details. The Lagrangian of \eqref{LSV} was implemented as a UFO model~\cite{Degrande:2011ua} by using {\sc FeynRules\,2.3.32}~\cite{Alloul:2013bka}.

\begin{figure}[t]
\includegraphics[width=0.5\textwidth]{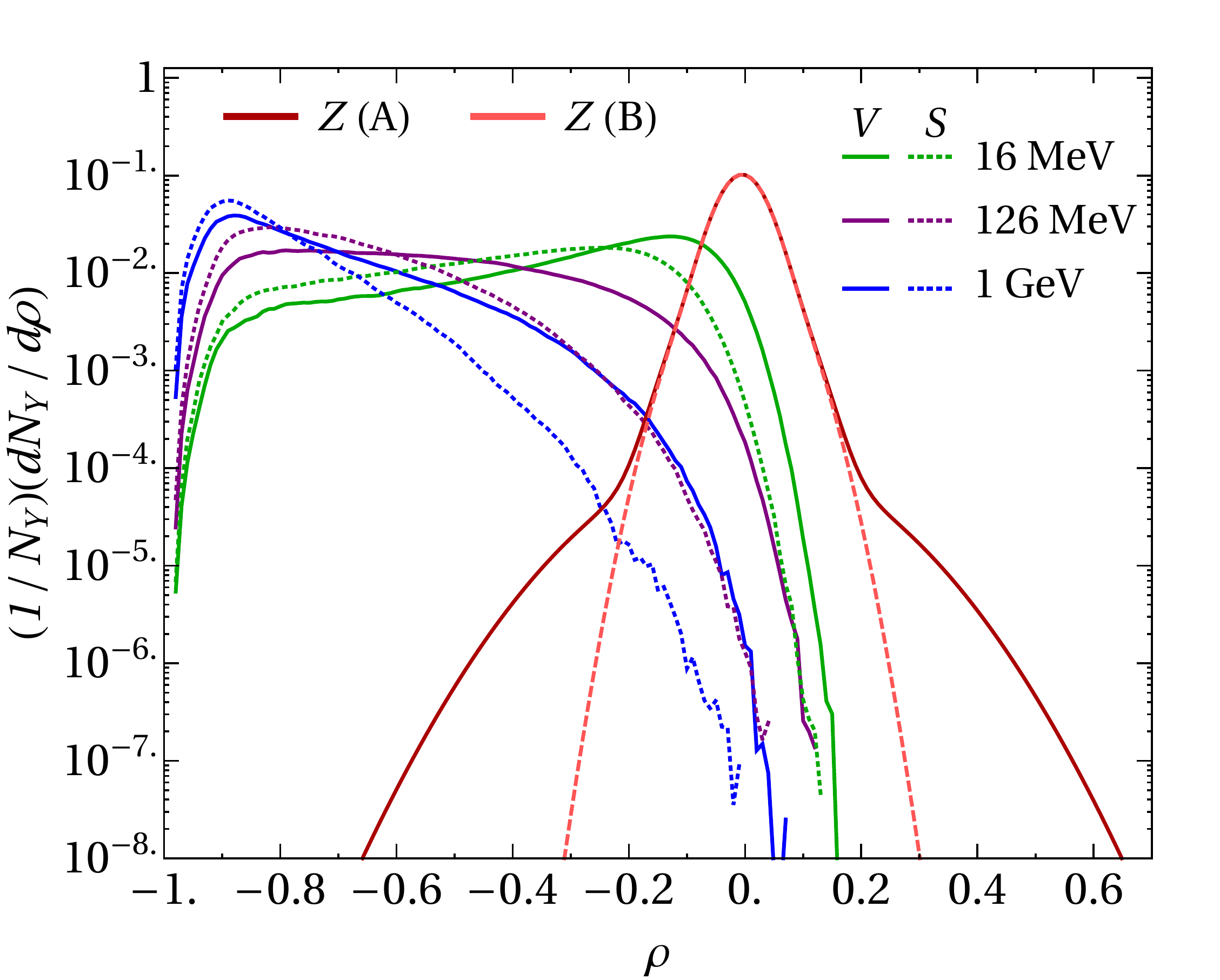}
\caption{The $\rho$ distribution, $(1/N_Y)\, dN_Y/d\rho$, for the SM (background) and MFC, i.e. $Y={\rm SM},\, S, \, V$, at different MFC masses.
} 
\label{fig:rho}
\end{figure}

The signal $\rho$ distributions $d\NMFC/d\rho$ are plotted in Fig.~\ref{fig:rho} for several MFC mass benchmarks. 
We validate our MC results by comparing them to the Weizs\"{a}cker-Williams~(WW) approximation~\cite{Kim:1973he,Tsai:1973py,Bjorken:2009mm}, and find $\cO(1)$ agreement, similar to Refs.~\cite{Liu:2016mqv,Liu:2017htz}.
%%%%%%%%%%%%%%%%%%%%%%%%%%%%%%%%%%%%%%%%%%%%%%%%
%%%%%%%%%%%%%%%%%%%%%%%%%%%%%%%%%%%%%%%%%%%%%%%%

Finally, combining the above, we estimate that $\NMFC\sim g^2_X(10^8 ,\, 10^7 ,\,10^5)$ for $m_X=16,\, 126,\, 1000\,$MeV, respectively, with $X=S$ or $V$.  
Thus, for an $\cO(1)$ acceptance, and assuming negligible background, we predict that this analysis is sensitive to $g_X$ in the range $10^{-4}-10^{-2}$. 

\subsecvspace

%%%%%%%%%%%%%%%%%%%%%%%%%%%%%%%%%%%%%%%%%%%%%%%%
\textit{Backgrounds} \newline
%%%%%%%%%%%%%%%%%%%%%%%%%%%%%%%%%%%%%%%%%%%%%%%%
The dominant expected background is due to true muons with mis-measured momenta. Motivated by ATLAS results in Ref.~\cite{Aad:2016jkr}, we model the momentum mis-measurement with a three Gaussian resolution function, centered at 0. 
The resolution as a function of $\rho$ is given by
\begin{align}
	\cR(\rho)
=	\frac{1}{\NSM}\frac{d\NSM}{d\rho} 
= 	\sum_{i=1,2,3}\frac{c_i}{\sqrt{2\pi\sigma^2_i}} e^{-\frac{\rho^2}{2\sigma^2_i}} \, ,
\end{align}
where $\sigma_1=0.035$, $\sigma_2=0.057$, $\sigma_3=0.15$, $c_1=0.75$ and $c_2=0.25$. 
The relative fraction of the third Gaussian is hard to model with the currently available public data. We therefore consider two representative cases 
\begin{align}
	c_3^{\rm A} = 5 \times 10^{-3} \, , \qquad 
	c_3^{\rm B} =  0\, . 
\end{align}
This resolution function includes hard photon radiation and possible subsequent hadronization effects mentioned in~\cite{Kahn:2018cqs}. 
The expected background $\rho$ distributions, $\cR$, for each of the above cases are plotted in Fig.~\ref{fig:rho}. 

In addition to true muons with mis-measured momenta, there can also be backgrounds from charged pion and kaon decays in-flight. If these decays happen in or near the calorimeter, they can mimic the MFC signal.
From publicly available information, we expect that the requirement of di-muon invariant mass to be within 5\,GeV of the $Z$ mass reduces these backgrounds to 5:10000 of the $Z \rightarrow \mu \mu$ sample.  
These backgrounds can be further rejected using an analysis of kink-in-tracks, and calorimeter information. 
Additionally, since muons from pions decaying in-flight have a minimal momentum fraction of 0.57 of the pion momentum, their contribution at $\rho$ close to $-1$ is small.
We estimate that the in-flight decay background can be rejected to a level between $10^{-7}$ and $10^{-8}$, and it is expected to be sub-dominant.
The $\rho$ distributions and overall normalization for pion and kaon decays can be extracted from a control sample with a same-charge requirement on the tag and probe particles.

\subsecvspace

%%%%%%%%%%%%%%%%%%%%%%%%%%%%%%%%%%%%%%%%%%%%%%%%
\textit{Reach for MFC} \newline
%%%%%%%%%%%%%%%%%%%%%%%%%%%%%%%%%%%%%%%%%%%%%%%%
We estimate the sensitivity of the proposed analysis to probe MFCs by a $d\NMFC/d\rho$ line shape analysis.
From the binned $\rho$ distribution, we construct a likelihood function, $L(g_X, m_X)$, and assume that the number of observed events is equal to the expected background events per bin.   
%%%%%%%%%%%%%%%%%%%%%%%%%%%%%%%%%%%%%%
\begin{figure*}[t]
\includegraphics[width=0.5\textwidth]{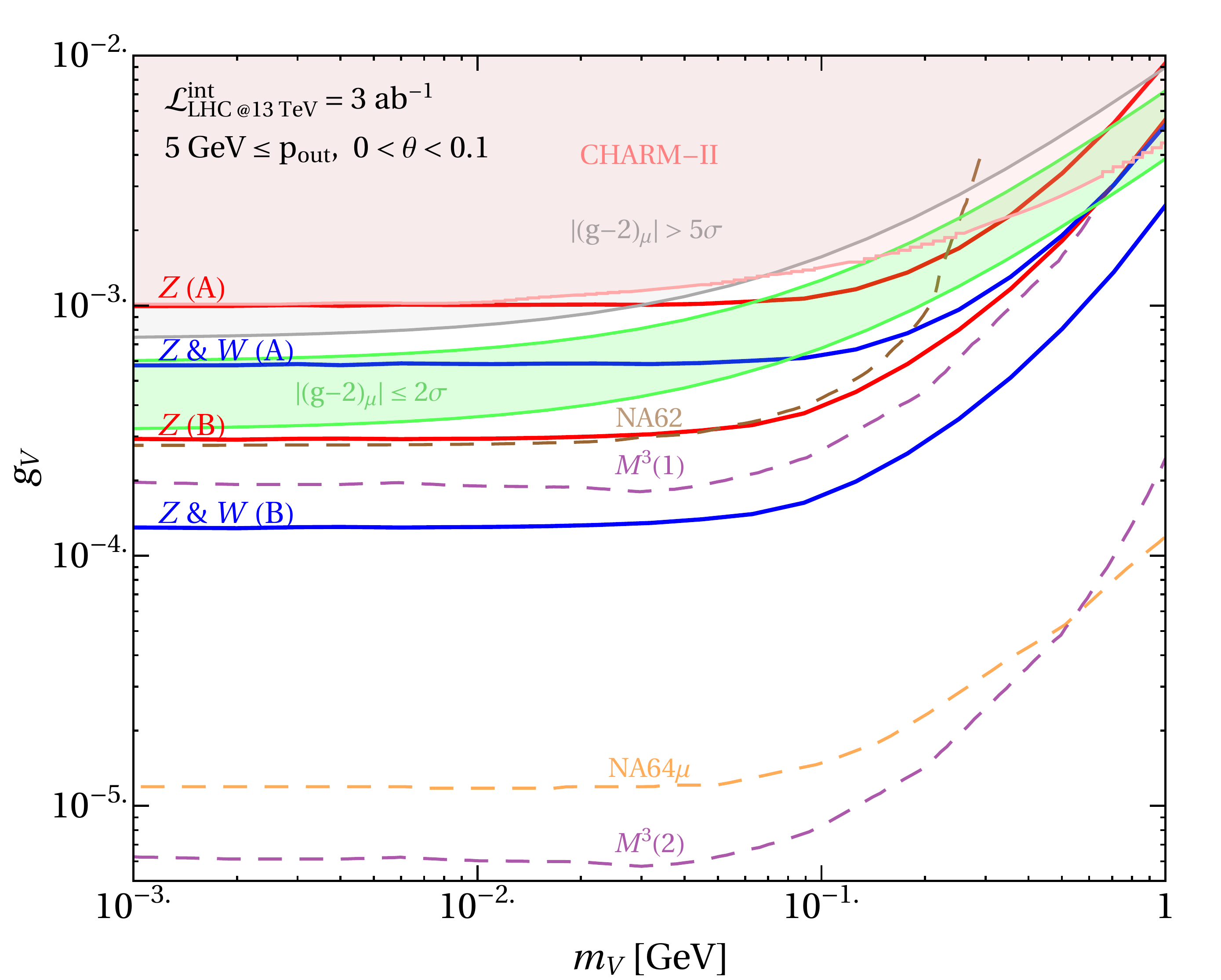}~
\includegraphics[width=0.5\textwidth]{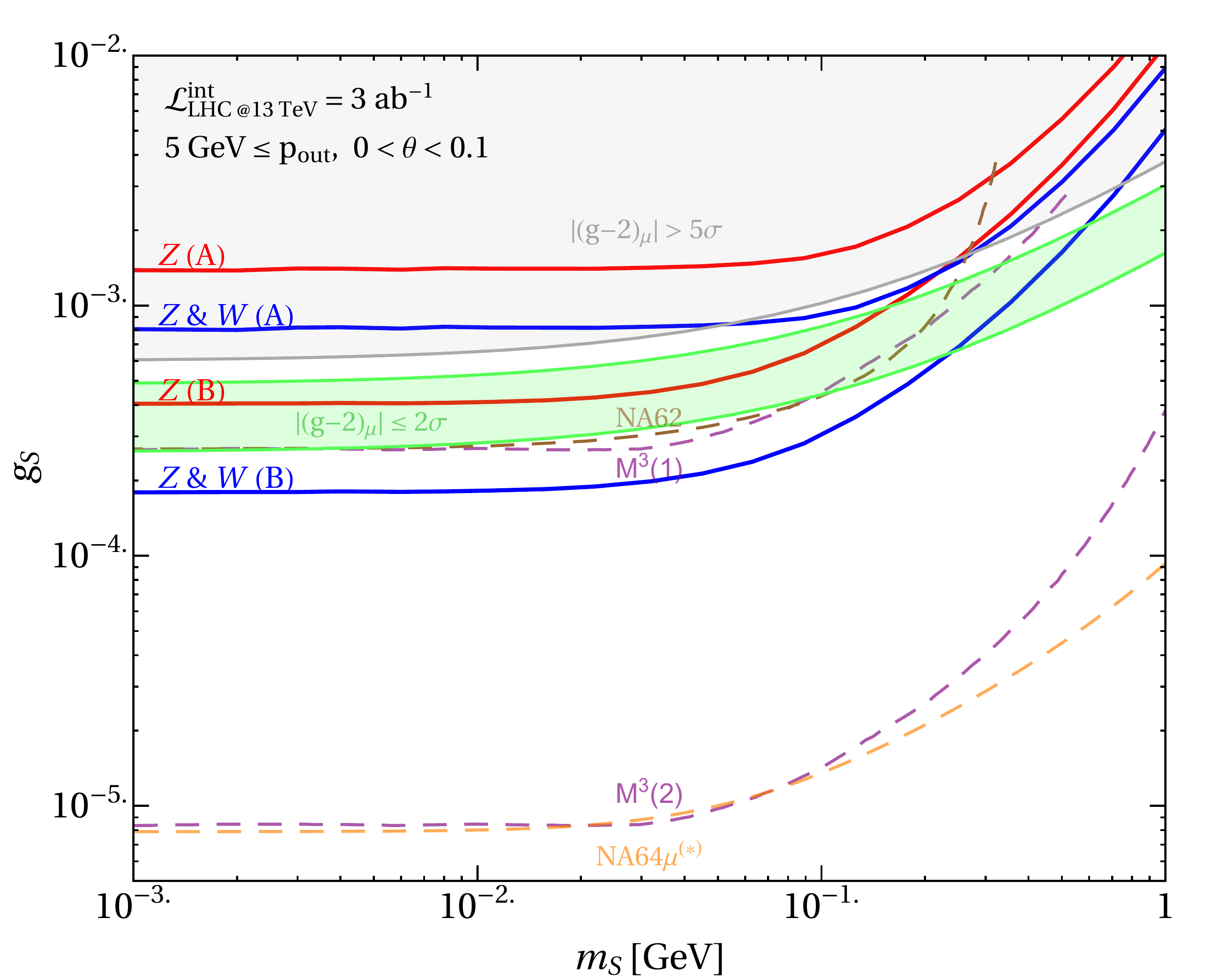}
\caption{The projections of the proposed ATLAS fixed-target like analysis to probe MFC at the HL-LHC comparing to current constraints from $(g-2)_\mu$~\cite{Bennett:2006fi,Beringer:1900zz} and CHARM-II~\cite{Geiregat:1990gz,Altmannshofer:2014pba} as well as to the projection of \mmm(1)(\mmm(2))~\cite{Kahn:2018cqs} with $10^{10}(10^{13})~\mu$ on-Target, NA62~\cite{Krnjaic:2019rsv} with $10^{13}~K^+$, and NA64$_\mu$~\cite{Chen:2018vkr,Gninenko:2019qiv} with $5\times10^{12}~\mu$ on-Target.
Left: vector mediator; right: scalar mediator. 
}
\label{fig:sensitivity_0p1}
\end{figure*}
%%%%%%%%%%%%%%%%%%%%%%%%%%%%%%%%%%%%%%
For a given $m_X$, we estimate the expected 95\,\% confidence level upper bound on $g_X$ for each of the above background scenarios and for two cases (i)~muons only from $Z$ decays; and (ii)~muons from both $Z$ and $W$ decays.

The projections are plotted in \figref{sensitivity_0p1} and are compared to the present bounds from $(g-2)_\mu$~\cite{Bennett:2006fi,Beringer:1900zz} and CHARM-II~\cite{Geiregat:1990gz,Altmannshofer:2014pba} as well as to the the projections of \mmm~\cite{Kahn:2018cqs}, NA62~\cite{Krnjaic:2019rsv} and NA64$_\mu$~\cite{Chen:2018vkr,Gninenko:2019qiv}.
We can see that the reach for $m_X\to 0$ can be at the level of $g_X\sim 10^{-3}-{\rm few} \times 10^{-4}$ (depending on the background model) by using only muons from $Z$. 
For the case of a combined analysis of $Z$ and $W$ muons, the sensitivity can reach the $g_X \sim 10^{-4}$ level.
This part of the parameter-space is relevant for the $(g-2)_\mu$ anomaly, to thermal freeze out dark matter scenarios, see {\it e.g.}~\cite{Kahn:2018cqs}, and comparable to other proposals such as \mmm~phase~1. 
Moreover, in the case that $X$ is a $L_\mu-L_\tau$ gauge boson, our proposal probes a $m_X-g_X$ parameter-space which may contribute to $N_{\rm eff}$ and possibly reduce the Hubble parameter tension~\cite{Escudero:2019gzq}. Finally, it is worth noting that while our sensitivity projections are for the ultimate HL-LHC dataset of $3~\iab$, even with the current dataset ($\sim 150~\ifb$) or the expected Run-3 dataset ($\sim 300~\ifb$), it should already be possible to probe at least a portion of these interesting regions of parameter-space.

\secvspace

%%%%%%%%%%%%%%%%%%%%%%%%%%%%%%%%%%%%%%%%%%%%%%%%
\textbf{\textit{Summary}} \newline
%%%%%%%%%%%%%%%%%%%%%%%%%%%%%%%%%%%%%%%%%%%%%%%%
We propose a search for NP using the large sample of muons produced at LHC collisions, and the ATLAS detector as a fixed-target experiment sensitive to missing muon momentum signatures.
In the proposed analysis, the calorimeter serves as a target for muons, and the muon momentum measurements before and after it are compared.

We focus on the possibility that a Muonic Force Carrier  is produced in the muon-target interaction and subsequently escapes the detector or decays invisibly. 
The detector signature corresponding to this scenario is of an unaccounted loss of muon momentum in the calorimeter. 
The expected sensitivity of the ATLAS experiment to this signature using muons from $Z$ and $W$ decays is comparable to other proposed experiments,
and overlaps the parameter-space that can explain the observations in the anomalous magnetic dipole moment of the muon.

%%%%%%%%%%%%%%%%%%%%%%%%%%%%%%%%%%%%%%%%%%%%%%%%%%%%%%%%%%%%%%%%
\begin{acknowledgments}
We thank 
John Paul Chou,
Yuri Gershtein,
Yonatan Kahn,
Gordan Krnjaic,
Scott Thomas,
and Yi-Ming Zhong
for useful discussions.
We thank 
Brian Batell,
Yonatan Kahn,
Gordan Krnjaic,
and Jesse Thaler for comments on the manuscript. 
IG and DS are supported by DOE grant DE-SC0010008.
EK is supported by ISF-8383/252 and ISF-1638/18 grants.
ST is supported by ISF-2181/15 and by I-CORE-1937/12 grants.
\end{acknowledgments}
%%%%%%%%%%%%%%%%%%%%%%%%%%%%%%%%%%%%%%%%%%%%%%%%%%%%%%%%%%%%%%%%

%%%%%%%%%%%%%%%%%%%%%%%%%%%%%%%%%%%%%%%%%%%%%%%%%%%%%%%%%%%%%%%%
\twocolumngrid
\vspace{-8pt}
\section*{References}
\vspace{-10pt}
\def\bibsection{}
\bibliographystyle{utphys}
\bibliography{muonic_forces_LHC}
%%%%%%%%%%%%%%%%%%%%%%%%%%%%%%%%%%%%%%%%%%%%%%%%%%%%%%%%%%%%%%%%

%%%%%%%%%%%%%%%%%%%%%%%%%%%%%%%%%%%%%%%%%%%%%%%%%%%
\input{muonic_forces_LHC_supp_arXiv_v1}

%%%%%%%%%%%%%%%%%%%%%%%%%%%%%%%%%%%%%%%%%%%%%%%%%%%

\end{document}

%% file: muonic_forces_LHC_supp_arXiv_v1.tex
\clearpage
\newpage
\maketitle
\onecolumngrid

\begin{center}
\textbf{\large Searching for muonic forces with the ATLAS Detector} \\
\vspace{0.05in}
\textit{\large Supplemental Material}\\
\vspace{0.05in}
{Iftah Galon, Enrique Kajamovitz, David Shih, Yotam Soreq and Shlomit Tarem}
\end{center}

\onecolumngrid
\setcounter{equation}{0}
\setcounter{figure}{0}
\setcounter{table}{0}
\setcounter{section}{0}
\setcounter{page}{1}
\makeatletter
\renewcommand{\theequation}{S\arabic{equation}}
\renewcommand{\thefigure}{S\arabic{figure}}
\renewcommand{\thetable}{S\arabic{table}}

%%%%%%%%%%%%%%%%%%%%%%%%%%%%%%%%%%%%%%%%%%%%%%%%
\section{Details of the signal yields calculation}
%%%%%%%%%%%%%%%%%%%%%%%%%%%%%%%%%%%%%%%%%%%%%%%%

Here we give a detailed derivation of the MFC signal production rates. 
We follow the schematics in \figref{illust}. 
A muon originating from the IP with momentum $\pin$ and direction $\eta$ interacts at a point $x$ in the material target of length $\ell_T(\eta)$, and produces a MFC, and an outgoing muon of momentum $\pout$ which travels in an angle $\theta$ relative to the incoming muon direction. 
The event yield is given by \eqref{yield},
\begin{align}
	\NMFC
&=\,	\LFT \,  \sigma_T
	 \int \! dp_{\rm in} \int_{\eta} d\eta \cA  \times  \cP_{\rm in}(\pin, \eta) \,  \, ,
\end{align}
where $\cA$ is the acceptance factor and it is a function of $\pin$ and $\eta$. 
Moreover, one can see $\cA$ as the probability for an incoming muon to scatter off the target and pass the ID-MS muon matching requirement for ME muons:
\begin{align}
	\label{eq:cA}	
	\cA
=	 \int_{0}^{\ell_T(\eta)} \! \frac{dx}{\ell_T(\eta)}
	\int_{\pout^{\rm min}}^{\pin} d\pout \int_{0}^{\theta^{max}(x,\eta)}\,d\theta \,
	\cP_{\rm out}(\pout,\theta\, | \, \pin) \,  ,
\end{align}
where $\theta^{max}(x,\eta)$ accounts for the ID-MS matching requirement and $\pout^{\rm min} = 5\,\GeV$ is taken as a minimal outgoing muon momentum in order to avoid resolution degradation in the MS measurement. 
The outgoing muon momentum distributions are given by:
\begin{align}
	\label{eq:Psig}
	\cP_{\rm out}(\pout,\theta \,|\, \pin ) 
	\equiv 
	\frac{1}{\sigma_T}{d^2\sigma_T\over d\pout d\theta} \, ,
\end{align}
where $\sigma_T$ is the fixed-target MFC production cross section, $\mu \, T \to \mu \, T  \, X $, see below for details. 

Note that in principle there could be additional acceptance or efficiency factors here to take into account more details of the ID-MS matching, other reconstruction or isolation efficiencies, and geometric acceptance factors having to do with details of the detector. We ignore these issues for simplicity and treat the efficiency factor as a constant over the fiducial range.

With these assumptions, the $\eta$ integral depends only on the initial process and can be performed once and for all, and the $x$ integral is trivial. 
Then \eqref{yield} reduces to
\begin{align}
	\label{eq:yieldapprox}
	\NMFC 
	&\approx\,
	\LFT\sigma_T  \int \! d\pin  \,  \cP_{\rm fid}(\pin) 
	\int_{\pout^{\rm min}}^{\pin} d\pout \int_{0}^{0.1}\,d\theta \,
	\cP_{\rm out}(\pout,\theta\, | \, \pin) \, ,
\end{align}
where we have introduced  the fiducial momentum distribution integrated over the ATLAS barrel region
\begin{align}
	\cP_{\rm fid}(\pin) 
	\equiv 
	\int_{\eta\in {\rm barrel}} d\eta\,\, \cP_{\rm in}(\pin,\eta) \, .
\end{align}
Finally, we write this in terms of the discriminating variable $\rho$ of \eqref{rho}, 
\begin{align}
	\label{eq:dNdrho}
	\frac{d\NMFC}{d\rho}
	\approx	
	\LFT\sigma_T  \int \! d\pin  \,  \cP_{\rm fid}(\pin) 
	\int_{0}^{0.1}\,d\theta \,
	\left.  \pin \cP_{\rm out}(\pout,\theta\, | \, \pin) \right|_{\pout=\pin(1+\rho)} \, ,	
\end{align}
where we consider only $\pout>\pout^{\min}=5\,\GeV\,$.

%%%%%%%%%%%%%%%%%%%%%%%%%%%%%%%%%%%%%%%%%%%%%%%%
\section{Calculation of $\mu\, T\to\mu \, T\, X$}
\label{sec:MFCxsect}
%%%%%%%%%%%%%%%%%%%%%%%%%%%%%%%%%%%%%%%%%%%%%%%%

Here we give the derivation of the $\mu\, T\to\mu \, T\, X$ cross section, where we closely follow Refs.~\cite{Liu:2016mqv,Liu:2017htz}.
We consider the following process
\begin{align}
	\label{eq:process}
	\mu (\pin) + T(P_i) \to \mu(\pout) + T(P_f) + X(k) \, , 
\end{align}
where $X=S$ or $V$ and the target ($T$) mass is $m_A$. 
In the lab-frame we use the following parameterization for the different momenta
\begin{align}
	\pin^\mu = \left( \Ein, 0, 0, \pin \right) \, , \qquad
	P_i^\mu = \left( m_A, 0, 0, 0 \right) \, , \qquad
	\pout^\mu = \left( \Eout, \pout s_\theta, 0, \pout c_\theta \right) \, ,
\end{align}
In addition, we define the following kinematical variables
\begin{align}
	P \equiv P_i + P_f \, , \qquad
	q \equiv P_i - P_f = \left( q_0, \vec{q} \right)  \, , \qquad 
	V \equiv \pout - \pin = \left( V_0 , \vec{V} \right) \, .
\end{align}

The differential cross section of the process in \eqref{process} is given by
\begin{align}
	d\sigma_T 
= 	\frac{1}{4 \pin m_A} \abs{\overline{\cM}^X_{2\to3}}^2 \,(2\pi)^4
	\delta^{(4)}(\pin+P_i - \pout - P_f - k)
	\frac{d^3 \pout}{(2\pi)^3 2 \Eout}
	\frac{d^3 P_f}{(2\pi)^3 2E_f}
	\frac{d^3 k}{(2\pi)^3 2E_k} \, ,
\end{align}
where $\abs{\overline{\cM}^X_{2\to3}}^2$ is the squared matrix element averaged over the incoming spins and summed over the outgoing spins.
Next, we perform the trivial integration over $d^3k$ using the $\delta^{(3)}()$-function. In order to integrate over the remaining $\delta()$-function, we
use $d^3P_f = d^3q$, and choose to describe $\vec q$ with respect to $\vec V$, i.e.\  $d^3q=Q^2 dQ\, dc_{\theta_{\vec V\vec q}} \,d\phi_q$ with the relative angle $\theta_{\vec V\vec q}$ between the vectors.
Then we fix $\theta_{\vec V\vec q}$ with the remaining $\delta()$-function to be 
\begin{align}
	c_{\theta_{\vec V\vec q}} &= \frac{m_X^2 + Q^2 + |\vec V|^2 - (V_0 - q_0)^2}{2 Q|\vec V|} \, .
\end{align}
Finally, defining $t\equiv -q^2 $, and integrating over the azimuthal angle of the vector $\pout$, we find
\begin{align}
	\label{eq:double_diff_xsect_full}
	\frac{d\sigma_T}{d\pout d c_\theta} 
= 	\frac{\pout^2}{64\pi^3 \pin \Eout |\vec V|}  \int_0^{2\pi} \frac{d\phi_q}{2\pi}
	\int_{t_{\rm min}}^{t_{\rm max}} \frac{dt}{8m_A^2} \abs{\overline{\cM}_{\,2\to3}}^2  \, ,
\end{align}
with 
\begin{align}
	t_{\rm max/min} 
=& 	2m_A \left( \sqrt{m^2_A + Q^2_{\rm max/min}} - m_A \right) \, , \\
	Q_{\rm{max} /\rm{min}} 
=&	\frac{((V_0 - m_A)^2 - |\vec V|^2 + m_A^2 - m_{X}^2)|\vec V| \pm
	(m_A - V_0)\lambda^{1/2}\big((V_0 - m_A)^2 - |\vec V|^2,m_A^2,m_{X}^2\big)}
	{2((V_0 - m_A)^2 - |\vec V|^2)} \, ,
	\label{eq:Q_def}
\end{align}
and $\lambda(x,y,z) = x^2 +y^2 + z^2 - 2(xy + xz + yz)\,$. 
The $(\pout, c_\theta)$ phase-space domain in \eqref{double_diff_xsect_full} is determined from \eqref{Q_def} by the requirement $\lambda\big((V_0 - m_A)^2 - |\vec V|^2,m_A^2,m_{X}^2\big)\ge 0$.
The $d\sigma_T/d\rho \, dc_\theta$ can be easily derived from \eqref{double_diff_xsect_full} by simple change of variable. 

The spin averaged/summed squared matrix elements for scalar, $X=S$, is
\begin{align}
	\abs{\overline{\cM}^S_{\,2\to3}}^2
=	g^2_S e^4 \frac{F(t)^2}{t^2}  \Bigg(
& 	(m_S^2 - 4m_\mu^2)
	\frac{(\tilde  s+ \tilde u)^2}{\tilde  s^2 \tilde u^2}
	\left( t(4m_A^2 + t) - 4\left(\frac{ (\pin \cdot P) \tilde u + (\pout\cdot P) \tilde s)}{(\tilde s + \tilde u)}\right)^2 \right)
	\nn\\
& + \frac{1}{\tilde s \tilde u} \left((\tilde s + \tilde u)^2(4m_A^2+t) - 4(k\cdot P)^2 t\right) \Bigg) \, ,
\end{align}
and for vector, $X=V$, is
\begin{align}
	\abs{\overline{\cM}^V_{\,2\to3}}^2
=	2g^2_V e^4 \frac{F(t)^2}{t^2} \Bigg(
&	\frac{ (m_V^2 + 2m_\mu^2 )  }{\tilde  s^2 \tilde u^2}\left(
	(4m_A^2 + t) t (\tilde{s}^2+\tilde{u}^2) -4\left( (\pin\cdot P) \tilde u + (\pout\cdot P) \tilde s \right)^2 \right) \nn\\
& 	-\frac{1}{\tilde s \tilde u} \Big( 4\left( (\pin\cdot P)^2 + (\pout\cdot P)^2 \right)t  \nn\\ 
&	- (4m_A^2+t) \left( (\tilde s + t)^2 + (\tilde u + t)^2 + (2m_\mu^2-m_V^2)t\right)\Big) \Bigg) \, .
\end{align}
In the above we use the following definitions 
\begin{align}
	\tilde{s} \equiv (\pout + k) ^2 -m^2_\mu = (\pin+q)^2 - m_\mu^2 \, , \qquad
	\tilde{u} \equiv (\pin - k)^2 - m^2_\mu = (\pout-q)^2 - m_\mu^2 \, .
\end{align}
The remaining scalar products are evaluated using
\begin{align}
	c_{\theta_{\vec{\pin} \vec q}} &= \frac{-(\pout s_\theta)(s_{\theta_{\vec V\vec q}}\,c_{\phi_q}) + (\pout c_\theta - \pin)c_{\theta_{\vec V\vec q}}}{|\vec V|} \, .
\end{align}
Finally, the form factors $F(t)$ for the atomic/nuclear targets are given by 
\begin{equation}
F(t) = \sqrt{G^{\rm{el}}_2(t) +G^{\rm{inel}}_2(t)}
\end{equation}
 where
the combined atomic and nuclear $G_2$ form-factors for elastic and inelastic processes are given by~\cite{Tsai:1973py,Kim:1973he,Bjorken:2009mm,Chen:2017awl}:
\begin{align}
	G^{\rm{el}}_2(t) 
&=  \left(\frac{a^2 t}{1+a^2 t}\right)^2 Z^2(1 + t/d)^{-2}\, , \\
	G^{\rm{inel}}_2(t) 
&= 	\left(\frac{a'^2 t}{1+a'^2 t}\right)^2 Z \left(\frac{1+t(\mu_p^2-1)/(4m_p^2)}{\left(1+t/(0.71~\GeV^2)\right)^4} \right)^2 \, , 
\end{align}
with the parameters
\begin{align}
	a
=	111\,Z^{-1/3}/m_e \, , \qquad
	d
=	0.164~\GeV^{2}\, A^{-2/3} \, , \qquad
	a' 
= 	773\, Z^{-2/3}/m_e \, , \qquad
	\mu_p 
= 	2.79 \, ,
\end{align}
where $m_e = 0.511\,\MeV$ is the electron mass, and $A,Z$ are the mass and atomic numbers of the target.
The virtuality $t\equiv|q^2|$ is defined using $q$, the four-momentum of the exchanged virtual photon.